\documentclass[a4paper,11pt]{article}
\usepackage{pos}
\usepackage{xfrac}

\usepackage{sidecap}
\sidecaptionvpos{figure}{t}

\title{Exploring the anisotropic HISQ (aHISQ) action}

\author[a,b]{Alexei Bazavov}
\author*[b]{Yannis Trimis}
\author[c]{Johannes H. Weber}

\affiliation[a]{Department of Computational Mathematics,
	Science and Engineering,\\
	Michigan State University, East Lansing, MI 48824, USA}

\affiliation[b]{Department of Physics and Astronomy,\\
	Michigan State University, East Lansing, MI 48824, USA}

\affiliation[c]{Institut f\"{u}r Physik \& IRIS Adlershof,\\ Humboldt-Universit\"{a}t zu Berlin D-12489 Berlin, Germany}

\emailAdd{trimisio@msu.edu}

\abstract{The fate of heavy quarkonia states in quark-gluon plasma is encoded in the temperature dependence of their spectral functions.
Reconstruction of spectral functions from Euclidean lattice correlators is an ill-posed problem.
Despite a variety of techniques developed recently, many questions remain unresolved.
It is known that the situation may be improved using anisotropic ensembles that provide finer resolution in the temporal direction.
To date, the effort focused on Wilson fermions.
We report on our first study with anisotropic improved staggered quarks.
To compute the spectrum of the anisotropic Highly Improved Staggered Quarks (aHISQ) we generated a library of anisotropic pure gauge ensembles.
We discuss the gauge anisotropy tuning that is performed with the Wilson and Symanzik gradient flow, as well as tuning of the strange quark mass and quark anisotropy with aHISQ, using spectrum measurements on quenched ensembles.
Finally, we discuss the impact of anisotropy on pion taste splittings for aHISQ.\\
\\
HU-EP-24/01-RTG
}

\FullConference{The 40th International Symposium on Lattice Field Theory (Lattice 2023)\\
July 31st - August 4th, 2023\\
Fermi National Accelerator Laboratory\\}

\begin{document}
\maketitle

\section{Introduction}

The melting pattern of heavy quarkonia and heavy open-flavor states in quark-gluon plasma is fully encoded in the temperature dependence of their spectral functions.
Reconstruction of the spectral functions from the Euclidean correlation functions computed in lattice QCD is an ill-posed problem which, despite of two decades~\cite{Asakawa:2000tr} of effort, has not yet been fully resolved.
While some progress has been made with the reconstruction methods (recent attempts include reconstruction with the Maximum Entropy Method, Backus-Gilbert method, a recently formulated Bayesian Reconstruction method~\cite{Burnier:2013nla}, Stochastic Optimization Method and machine learning techniques based on neural networks), it is of primary importance to produce gauge field ensembles that are specifically tuned for addressing the reconstruction problem.
Namely, one is looking for ensembles with (potentially) very high statistics and large temporal extents $N_\tau$. Isotropic ensembles with large $N_\tau$ are prohibitively expensive, however, making lattices anisotropic, \textit{i.e.}, $a_{\sigma}/a_{\tau}=\xi\sim 6 - 8$ alleviates that problem, where $a_{\sigma}$ ($a_{\tau}$) is the spatial (temporal) lattice spacing.
In terms of statistics, staggered fermions have proven to be the leading formulation for finite-temperature QCD with some ensembles reaching order million configurations.
Naive (\textit{i.e.}, unimproved) staggered fermions feature large discretization effects in the hadron spectrum, and early attempts with anisotropic naive staggered fermions~\cite{Nomura:2004qu,Levkova:2006gn} have not been further pursued.
However, development of improved actions such as HISQ~\cite{Follana:2006rc} allowed one to significantly suppress the discretization effects in the fermion sector.
We therefore explore an anisotropic variant of the HISQ action that we call anisotropic Highly Improved Staggered Quarks (aHISQ).

It is important to mention that anisotropic ensembles with dynamical fermions exist and are being used for the spectral function reconstruction problem by the FASTSUM collaboration, \textit{e.g.}, Ref.~\cite{FASTSUM:2022qzx}. However, they are based on the fixed scale approach with Wilson fermions. Wilson fermions are more computationally demanding and therefore are inherently more limited in statistics. The fixed scale approach, while being more economical for tuning, leads to decreasing the temporal extent $N_\tau$ with increasing temperature. 
In the temperature range relevant for the physics of quark-gluon plasma these ensembles at present have $N_\tau\le 36$~\cite{FASTSUM:2022qzx}. Such temporal extents are in the range that can be reached today on isotropic lattices with staggered fermions, \textit{e.g.}, Ref.~\cite{Bazavov:2023dci}, and thus one should be able to reach much larger $N_\tau$ with an anisotropic staggered fermion formulation.

\section{Tuning of the gauge anisotropy}

At this stage of the project we experiment with anisotropic pure gauge ensembles where we can test anisotropy tuning and study the aHISQ spectrum without the complications of dynamical fermions. For brevity, we skip the details of introducing the anisotropy in the gauge action, as the method is rather standard and discussed in detail, for instance, in Ref.~\cite{Burgers:1987mb}. The most relevant point to mention is that we use the tree-level Symanzik-improved gauge action (plaquette  + rectangle).

The gradient flow introduced in Ref.~\cite{Luscher:2010iy} has found many uses in the lattice community and turned out to be particularly useful for scale setting. The procedure was extended to include the anisotropy in Ref.~\cite{Borsanyi:2012zr}. The original proposal there introduced the flow anisotropy $\xi_{gf}$ which enters the flow equation (\textit{e.g.}, for the Wilson flow)
\begin{equation}
\frac{dU_\mu}{d t}= X_\mu(U) U_\mu,\,\,\,\,\,\,\,
X_\mu(x,t)= \mathcal{P}_A\left[ \sum_{\pm\nu\ne \mu} \rho_{\mu\nu}U_\nu(x,t) U_\mu(x+\nu,t) U^\dagger_\nu(x+\mu,t)U^\dagger_\mu(x,t) \right]
\end{equation}
through the weight factors $\rho_{\mu\nu}$ such that $\rho_{i4}=\xi_{gf}^2$ and
$\rho_{ij}=\rho_{4i}=1$. The observable (the action density) is split into the temporal and spatial parts
\begin{equation}
S_{\sigma\sigma}(t)=\frac{1}{4}\sum_{x,i\ne j} F_{ij}^2(x,t)\,,\,\,\,\,\,\,\,\,\,\,
S_{\sigma\tau}(t)=\frac{1}{2}\sum_{x,i}F_{i4}^2(x,t)
\end{equation}
and for the $w_0$ scale the ratio of derivatives is defined as
\begin{equation}
R_E=\left. \left[t \frac{d}{d t} t^2 \langle S_{\sigma\sigma}(t) \rangle\right]_{t=w_0^2}\middle/\left[t\frac{d}{d t} t^2 \langle S_{\sigma\tau}(t) \rangle\right]_{t=w_0^2}\right.\,.
\label{eq_RE}
\end{equation}
The set of equations for simultaneously determining the spatial lattice spacing and the renormalized gauge anisotropy for a given ensemble is then
\begin{equation}
\left[t\frac{d}{d t} t^2 \langle S_{\sigma\sigma}(t) \rangle\right]_{t=w_0^2} 
= 0.15\,,\,\,\,\,\,\,\,\,\,\,
R_E(\xi_{gf}^2)/\xi_{gf}^2 = 1.
\label{eq_ani1}
\end{equation}
The procedure encoded in Eq.~(\ref{eq_ani1}) requires running several flows per lattice to determine the point where the second equation is satisfied (by interpolation or extrapolation). The flow anisotropy at that point is the \textit{defined} gauge anisotropy of the ensemble and the $w_0$ at the same point \textit{defines} the spatial lattice spacing.

\begin{figure}
	\begin{center}
		\includegraphics[width=0.33\textwidth]{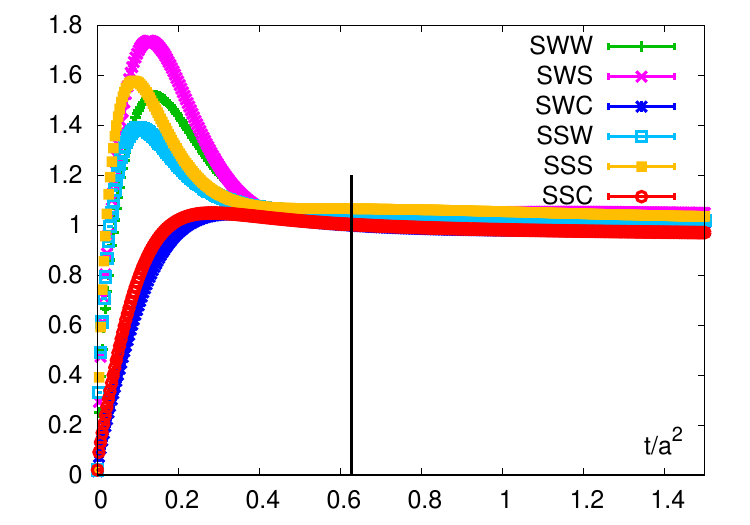}\hfill
		\includegraphics[width=0.33\textwidth]{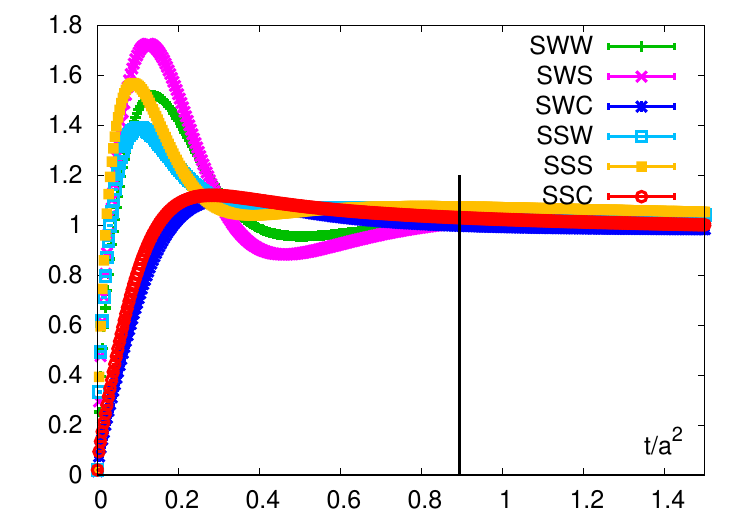}\hfill
		\includegraphics[width=0.33\textwidth]{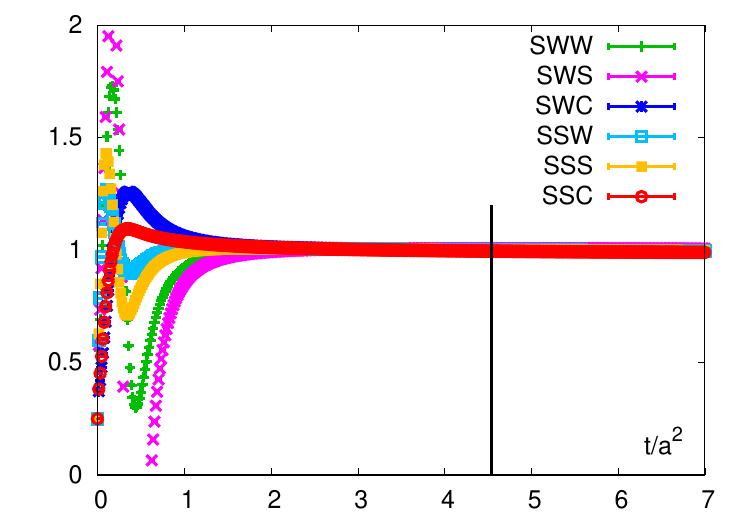}
		\caption{Evolution of the ratio $R_E$ which is used to define the renormalized gauge anisotropy, Eq.~(\ref{eq_RE}), with flow time for three pure gauge ensembles: $a_{\sigma}=0.22$~fm, $\xi=8$ (left), $a_{\sigma}=0.19$~fm, $\xi=8$ (middle), $a_{\sigma}=0.08$~fm, $\xi=2$ (right). The statistical errors are smaller than symbol sizes.\label{fig_flows}}
	\end{center}
\end{figure}
In Fig.~\ref{fig_flows} we show the evolution of the ratio $R_E$ with the flow time. We explore two types of flow (Wilson and Symanzik) and three discretization schemes for the observable (Wilson, Symanzik and clover). The gauge action is fixed to the tree-level Symanzik-improved action. The combination of the gauge action, flow action and observable is encoded in a three letter abbreviation, in that order. The vertical lines indicate the flow time $t=w_0^2$ in the SSC scheme. The ensemble in the left panel has the spatial lattice spacing $a_{\sigma}=0.22$~fm and the renormalized anisotropy $\xi=8$, in the middle panel $a_{\sigma}=0.19$~fm, $\xi=8$ and in the right panel $a_{\sigma}=0.08$~fm, $\xi=2$. We observe that the SSC and SWC combinations approach the plateau in a smoother way than the other combinations. This plays a role for tuning coarse ensembles since if the flow time $t=w_0^2$ does not correspond to a plateau but is small enough to probe early flow time artifacts, the flow tuning procedure would amplify the lattice artifacts. We conclude that with the renormalized anisotropy $\xi=8$ we can reach lattice spacings as coarse as $a_{\sigma}=0.22$~fm. This may allow for using the non-relativistic QCD (NRQCD) formalism for the charm quark. For the finer $a_{\sigma}=0.19$~fm lattice we observe that the early flow time artifacts may still be present around the time $t=w_0^2$ for some (SWW, SWS) combinations and SSC and SWC provide the most convenient choice. For a fine lattice such as the one ($a_{\sigma}=0.08$~fm) shown in the right panel (although at a smaller renormalized anisotropy $\xi=2$) of Fig.~\ref{fig_flows}, all combinations reach the plateau well before the flow time $t=w_0^2$. We conclude that the most robust choices for tuning, if one is interested in coarse ensembles, are the SSC and SWC combinations.

 Later, in Ref.~\cite{Borsanyi:2018srz} another variant was introduced that requires only one flow per lattice, where instead of tuning the ratio one directly matches the separately defined spatial and temporal $w_0$-scales:
\begin{equation}
\left[t\frac{d}{d\tau} t^2 \langle S_{\sigma\sigma}(t) \rangle\right]_{\tau=w_{0,\sigma}^2} = 0.15\,,\,\,\,\,\,\,\,\,\,\,
\left[t\frac{d}{d\tau} t^2 \langle S_{\sigma\tau}(t) \rangle\right]_{t=w_{0,\tau}^2} = 0.15\,.
\label{eq_tune1}
\end{equation}
In this case, the flow anisotropy is set directly to the target renormalized anisotropy and the condition that the target anisotropy has been achieved is
\begin{equation}
\frac{w_{0,\sigma}}{w_{0,\tau}}=1.
\label{eq_tune2}
\end{equation}
One still needs to scan (\textit{i.e.}, to generate ensembles) the two-dimensional parameter space of the bare gauge coupling $\beta$ and the bare gauge anisotropy $\xi_0$, but, unlike with the previous approach, one needs to run only one flow per lattice. 
We tested both approaches and, as Ref.~\cite{Borsanyi:2018srz}, found the second one to be more economical.

Initially, we have chosen the SSC combination as the preferred method for tuning the gauge anisotropy. On the coarse ensembles ($a_{\sigma}>0.16$~fm) we noticed that the bare anisotropy $\xi_0$ is numerically closer to the renormalized anisotropy $\xi$. However, it turned out that the SSC combination has a significant disadvantage. To explore it, we tuned several ensembles with the bare gauge coupling $\beta\equiv 10/g_0^2=6.9$, $7.0$, $7.1$ and $7.2$ and the renormalized anisotropy $\xi=2$ using all six available flow/observable combinations. (We used the second anisotropy tuning method, Eqs.~(\ref{eq_tune1}) and (\ref{eq_tune2}), in this case.)

\begin{SCfigure}
	\centering
	\includegraphics[width=0.6\textwidth]{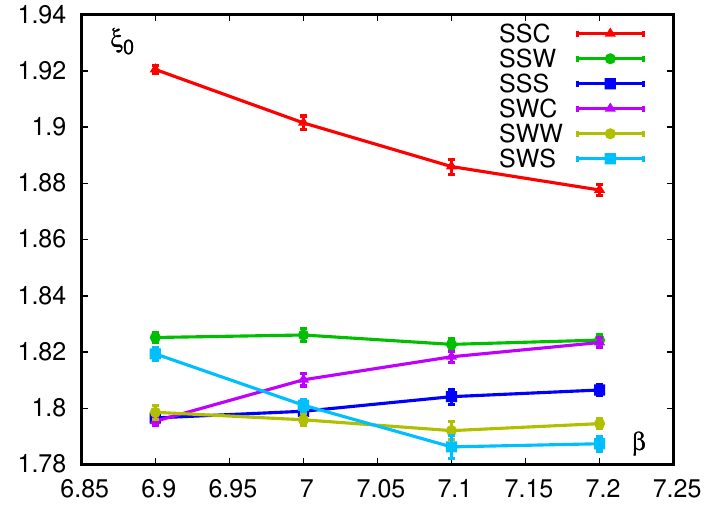}
	\caption{Dependence of the bare gauge anisotropy $\xi_0$ on the bare gauge coupling $\beta$ which corresponds to the same renormalized gauge anisotropy $\xi=2$ with six different combinations of the flow and observable.
		\label{fig_xi0_beta}
	}
\end{SCfigure}
In Fig.~\ref{fig_xi0_beta} we show the dependence of the bare gauge anisotropy $\xi_0$ on the bare gauge coupling $\beta$ for these ensembles. In other words, the lines in Fig.~\ref{fig_xi0_beta} are the lines of constant renormalized anisotropy (LCRA), $\xi=2$ in this particular case.
We expect that as $\beta\to\infty$ all lines should approach 2.
The SSC combination, although being closest to 2, is \textit{decreasing} with $\beta$.
This means that the dependence $\xi_0(\beta)$ along the SSC LCRA is non-monotonic, and at large enough $\beta$ it should reach a minimum and then start increasing.
This is an undesirable complication, since it means that the continuum extrapolation along the SSC LCRA is going to be non-monotonic and to truly reach the scaling regime we need to work at much larger $\beta$.
In contrast, SWC is monotonically increasing (in fact, the same SWC LCRA in a larger range of couplings is shown in Fig.~2 of Ref.~\cite{Borsanyi:2018srz}).
We therefore conclude that SWC is a preferable way of tuning the gauge anisotropy, especially if one needs to tune coarse ensembles.

\section{Tuning the fermion anisotropy}

We define the anisotropic HISQ (aHISQ) action as
\begin{align}
&S_{\text{aHISQ}}[{{\chi}}, \bar{{\chi}}]= 
\frac{a_{\sigma}^4}{\xi_0^f} \sum\limits_{n,m \in {\Lambda}}
		\bar{{\chi}}(n)
		\nonumber\\
		&\times\left\{
		\sum\limits_{\pm\mu=0}^{3} \frac{\eta_{\mu}(n)(\xi_0^f)^{\delta_{|\mu|0}}}{2a_{\sigma}}
		\Big[V^{\text{F}}_{\pm\mu}(n)\delta(n \pm\hat{\mu},m) 
		+V^{\text{L}}_{\pm\mu}(n)\delta(n \pm3\hat{\mu},m) \Big]
		~+~m_0 \delta(n,m)
		\right\}~
		{\chi}(m)
		\label{eq:SaHISQ}~,
\end{align}
where $\xi_0^f = a_{\sigma}/a_{\tau}$ is the bare fermion anisotropy that simply multiplies the same fat- or long-links ($V^{\text{F,L}}_{\mu}$) with the standard coefficients as for the isotropic HISQ action, and for the rest of the notation we refer the reader to Ref.~\cite{Follana:2006rc}. 
Note that since the smearing coefficients of the different staples are chosen in order to cancel the tree-level contribution from hard transverse momenta at the cutoff $\pi/a_{\sigma}$, the anisotropy cannot contribute to the relative staple weights. 
As can be seen from Eq.~\eqref{eq:SaHISQ}, the modification of the HISQ action is minimal, the smearing and reunitarization of the links are not affected and the fermion anisotropy $\xi_0^f$ enters at the outermost smearing level.

The physical fermion anisotropy is defined through the hadronic dispersion relation
\begin{equation}
  E^2(\xi^f,M^2,p^2) = M^2 + \frac{p^2}{(\xi^f)^2}
  \label{eq:disp}~
\end{equation}
for a correlation function in the anisotropic temporal direction. 
We simultaneously tune the bare strange quark mass in lattice units $a_{\sigma}m_s$ such that the fictitious Goldstone boson $\eta_{s\bar s}$ (taste pseudoscalar meson with only strange quarks) coincides with the prediction from chiral perturbation theory, \textit{i.e.}, $m_{\eta_{s\bar s}}= 686$ MeV, and the bare fermion anisotropy such that the renormalized fermion anisotropy agrees with the renormalized gauge anisotropy.
We then take the fixed ratio $m_s/m_l=5$ for the light quarks which results in the pion mass of about 300~MeV. 
We use hadron momenta up to $\sqrt{2}\cdot 2\pi/L$. 

\begin{SCfigure}
\centering
\hspace{-0.5cm}\includegraphics[width=0.6\textwidth]{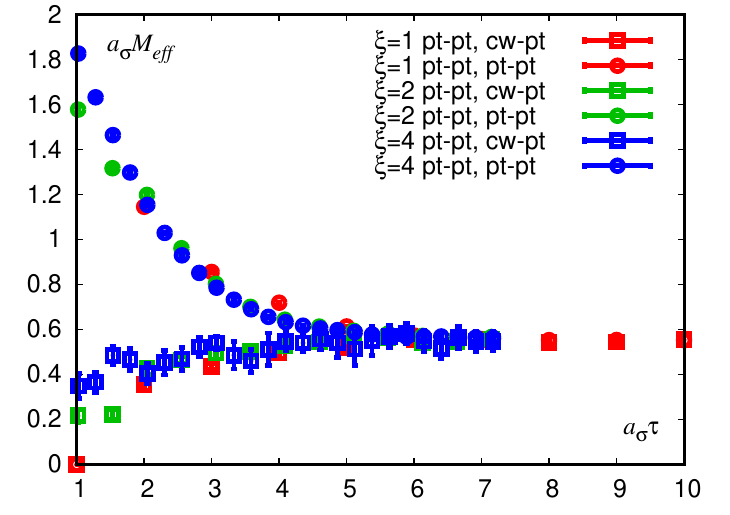}
\caption{Pseudoscalar meson effective mass extracted from correlators with different source-sink combinations. The effective mass as well as the temporal distance are measured in the units of the spatial lattice spacing to compare ensembles with different anisotropies, $\xi=1$, $2$ and $4$. 
\label{fig:meff}
}
\end{SCfigure}

For measuring the staggered meson spectrum we use 
point, random wall, or (corner) wall sources, and point or wall sinks. 
We use two source positions per configuration.
Since point sources excite all quark momenta, different momentum projections can be done at the sink for each pair of propagators with the same point source. 
Random wall sources select one specific momentum (usually $p=0$) that has to be injected into one of the two propagators at the source, to which one has to project at the sink, too. 
The extra cost is outweighed by their better signal-to-noise ratio, while the mean agrees within errors with point source correlators.
Due to their symmetry and thus strictly positive spectral weights correlators with point sink, and point or random wall source approach the effective mass plateau from above.
Correlators with (corner) wall at the source or the sink yield a better signal-to-noise ratio, too, but approach the plateau from below (permitting negative spectral weights due to their lack of symmetry). 
As before, explicit momentum injection is necessary for the wall source. 
Combining wall and point-type operators, the plateau can be identified even for a physically short time  direction, as shown in Fig.\ref{fig:meff}. 
The figure also shows that the overlap factors and energy splittings have at most a mild dependence on the anisotropy. 

\section{Measuring the pion taste mass splittings}

Taste is an unphysical degree of freedom of staggered quarks with an only approximate -- due to the roughness of the gauge fields -- internal symmetry for QCD at finite lattice spacing, \textit{i.e.}, $SO(4)$~\cite{Lee:1999zxa} in the isotropic case. 
Staggered hadron correlators always receive contributions from multiple taste combinations, which also encode different hadronic parities. 
The corresponding parity partner contribution includes due to the pole structure an alternating sign, and the hadron of interest may be either of the two.
At the tuned value of $m_l=m_s/5$ we measure the different pion tastes. 

\begin{figure}
	\includegraphics[width=\textwidth]{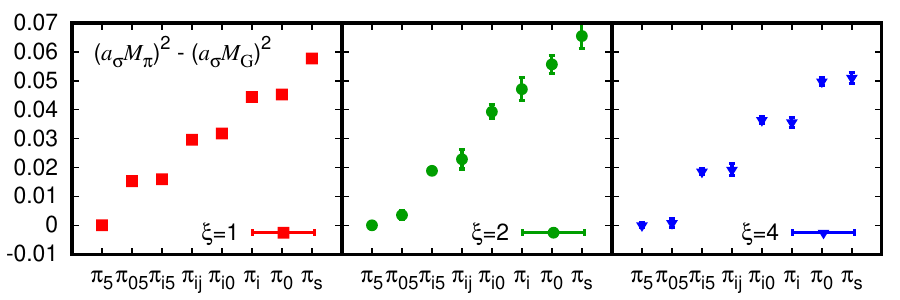}
	\caption{Quadratic mass splittings of the pion tastes from the Goldstone ($\gamma_5$ taste) pion on a quenched QCD ensemble with the spatial lattice spacing $a_\sigma\approx0.16$~fm and the renormalized gauge anisotropy $\xi=1$ (left), 2 (middle) and 4 (right).
		\label{fig:pisplit}
	}
\end{figure}

The well-known $SO(4)$ multiplet pattern $(1,4,6,4,1)$ is indeed observed for staggered pion tastes in the left panel of Fig.~\ref{fig:pisplit}. 
However, as the anisotropy is increased from 1 this $SO(4)$ pattern disappears, and a new symmetry pattern starts to form, as shown in the right panel of Fig.~\ref{fig:pisplit}. 
The emerging, duplicated pattern $(1,3,3,1)$ can be understood as being due to the increasingly smoother distribution of the temporal links, whereby the temporal taste generator $\gamma_{0}$, that could be realized for a staggered one-component field with a single temporal lattice step (and an appropriate phase) becomes increasingly similar to the identity operator in taste space (with an alternating sign in the temporal direction), 
which is exactly the expectation from the free theory. 
This partial taste symmetry restoration is a very strong hint that the full taste symmetry restoration in the approach to the continuum (also for isotropic staggered quarks) is a smooth process without discontinuities. 

\section{Conclusion}

We have presented first results on the pion spectrum with the anisotropic HISQ (aHISQ) action.
The aHISQ pion correlation functions were calculated on anisotropic pure gauge ensembles generated with the tree-level Symanzik-improved action and with the renormalized gauge anisotropies $\xi=1$, $2$ and $4$.
Working with pure gauge ensembles allows us to split the problem in two: initial tuning of the gauge coupling and anisotropy with subsequent independent tuning of the strange quark mass and fermion anisotropy.
With dynamical fermions all four parameters will need to be tuned simultaneously.
We have experimented with different gauge anisotropy tuning schemes. All of them employ the gradient flow but differ in what flow/observable combination is used and what quantities are used to define the anisotropy. For the pion taste multiplets we observe how the symmetry pattern changes with increasing the anisotropy. The change is consistent with expectations.

\section*{Acknowledgements}

A.B and Y.T.'s research is funded by the U.S. National Science Foundation under the awards 
PHY-1812332 and PHY-2309946. J.H.W.’s research is funded by the Deutsche Forschungsgemeinschaft (DFG, German Research
Foundation) -- Projektnummer 417533893/GRK2575 ``Rethinking Quantum Field Theory''. 

Computational resources used in this work were in part provided by the
Institute for Cyber-Enabled Research at Michigan State University and
the USQCD Collaboration, funded by the Office of Science of the U.S.\ Department of Energy.

\end{document}